\documentclass[aps,prd,twocolumn,superscriptaddress,nofootinbib,longbibliography]{revtex4-2}

\usepackage[T1]{fontenc}
\usepackage[utf8]{inputenc}
\usepackage{lmodern}
\usepackage{amsmath,amssymb,mathtools,bm}
\usepackage{microtype}
\usepackage{tikz}
\usetikzlibrary{calc,arrows.meta,positioning}
\usepackage{comment}
\usepackage[colorlinks=true,citecolor=blue,linkcolor=blue,urlcolor=blue]{hyperref}
\usepackage{mathrsfs}
\usepackage{braket}
\allowdisplaybreaks

\newcommand{\dd}{\mathrm{d}}
\newcommand{\iu}{\mathrm{i}}

\newcommand{\Id}{\mathsf 1}

\usepackage{amsthm}  

\newtheorem{proposition}{Proposition}
\newtheorem{theorem}{Theorem}

\theoremstyle{definition}

\theoremstyle{remark}

\definecolor{lime}{HTML}{A6CE39}
\DeclareRobustCommand{\orcidicon}{%
 \begin{tikzpicture}
 \draw[lime, fill=lime] (0,0) 
 circle [radius=0.16] 
 node[white] {{\fontfamily{qag}\selectfont \tiny ID}};
 \draw[white, fill=white] (-0.0625,0.095) 
 circle [radius=0.007];
 \end{tikzpicture}
 \hspace{-2mm}
}

\foreach \x in {A, ..., Z}{%
 \expandafter\xdef\csname orcid\x\endcsname{\noexpand\href{https://orcid.org/\csname orcidauthor\x\endcsname}{\noexpand\orcidicon}}
}

\newcommand\orcidDavid{{\href{https://orcid.org/0009-0006-8485-3140}{\orcidicon}}}
\newcommand\orcidVishnu{{\href{https://orcid.org/0000-0001-8523-4647}{\orcidicon}}}
\newcommand{\mpl}{\ell_{\rm P}}

\newcommand{\TrF}{\operatorname{Tr}_{F}}

\newcommand{\order}{\mathcal{O}}
\begin{document}

\title{Scalar Black-Holes in Gravity from Entropy}

\author{David S. Pereira\orcidDavid\!\!}
\email{djpereira@ciencias.ulisboa.pt}
\affiliation{Departamento de F\'{i}sica, Faculdade de Ci\^{e}ncias da Universidade de Lisboa, Campo Grande, Edif\'{i}cio C8, P-1749-016 Lisbon, Portugal}
\affiliation{Instituto de Astrof\'{i}sica e Ci\^{e}ncias do Espa\c{c}o, Faculdade de Ci\^{e}ncias da Universidade de Lisboa, Campo Grande, P-1749-016 Lisbon, Portugal}

\author{Udaykrishna Thattarampilly}
\email{uday7adat@gmail.com}
\affiliation{Center for Gravitation and Cosmology, College of Physical Science and Technology,\\
Yangzhou University, Yangzhou 225009, China}

\author{Vishnu Kakkat\orcidVishnu\!\!}
\email{vishnu.kakkat@nithecs.ac.za}
\affiliation{Institutionen f\"or teknikvetenskap och matematik, Lule\aa{} tekniska universitet, 971 87 Lule\aa{}, Sweden}
\affiliation{National Institute for Theoretical and Computational Sciences (NITheCS), Stellenbosch, South Africa}

\date{\today}

\begin{abstract}
We investigate static, spherically symmetric black holes in Gravity from Entropy (GfE), focusing on the sector in which the bosonic form-space field is a pure scalar. Starting from the full auxiliary-field equations, we show that the Dirac--K\"ahler matter equation contains both a scalar equation and a two-form consistency condition. The latter is automatically satisfied by a radial scalar. We derive an exact integral identity, valid for a general backreacted static spherical geometry, and prove that a regular scalar approaching zero at infinity must vanish if the sampled scalar and radial auxiliary-field eigenvalues are positive and the effective mass squared $m^2+\xi R$ is nonnegative. In particular, this excludes minimally coupled static scalar hair with $m^2\geq0$ on the regular identity-connected branch. We also present the perturbative GfE vacuum geometry and use its curvature to identify the region in which nonminimal coupling can evade the sign argument. A second, sharper estimate shows that making $m^2+\xi R$ negative is not by itself sufficient to produce a regular static zero mode: a finite interval of nonminimal couplings still admits no branch bifurcating from the vacuum solution.
\end{abstract}

\maketitle

\section{Introduction}\label{sec:intro}

Black holes occupy an important position in gravitational physics: they are at once the simplest nontrivial solutions of the Einstein equations and systems for which gravity, thermodynamics, and quantum theory meet. The identification of black-hole entropy with horizon area and the subsequent derivation of Hawking radiation established that horizon geometry carries thermodynamic information \cite{Bekenstein:1973ur,Hawking:1975vcx}. Related ideas motivate approaches in which aspects of the gravitational equations are understood in terms of entropy or local thermodynamic balance \cite{Jacobson:1995ab}. These developments make black holes natural laboratories for theories in which entropy enters not only through a property of a solution, but directly through the gravitational action.

The external fields that a stationary black hole can support provide another test of a gravitational theory. Classical no-hair arguments constrain regular scalar configurations subject to specified symmetry, asymptotic, and positivity assumptions \cite{Chase:1970omy,Bekenstein1995,Bekenstein1995,Mayo:1996mv,Heusler:1996jaf,Chrusciel:2012jk}. Their lesson is not that scalar hair is universally forbidden, but that its existence depends on the structure of the scalar equation and its boundary conditions. Reviews of asymptotically flat black holes with scalar hair emphasize the roles of symmetry inheritance and the sign of the relevant kinetic or potential terms can be found in~\cite{Herdeiro:2015waa}. In modified gravity, a small change in the curvature coupling can alter the hypotheses underlying a familiar no-hair theorem \cite{Graham:2014mda}.

Derivative interactions can modify the usual current-based exclusion arguments \cite{Hui:2012qt,Babichev:2013cya}; a rotating black hole can support a synchronized complex scalar whose stress tensor is stationary even though the field carries a harmonic time dependence \cite{Herdeiro:2014goa}; and curvature-dependent effective masses can trigger spontaneous scalarization \cite{Silva:2017uqg,Doneva:2017bvd}. These constructions do not contradict the classical theorems, because one or more assumptions have changed. For a new gravitational model, it is therefore essential to determine the actual matter equation before importing a result derived for an ordinary Klein--Gordon field.

Gravity from Entropy (GfE) is a particularly instructive example. In its foundational formulation, its bosonic degrees of freedom are organized on the direct sum of differential-form spaces $\Lambda^0\oplus\Lambda^1\oplus\Lambda^2$, and both matter and curvature contribute to a mixed endomorphism whose trace logarithm defines the action \cite{Bianconi:2024aju}. This formulation establishes GfE as an interesting theory to explore in terms of BH solutions as matter plays a fundamental role in its formulation and dynamics. Moreover, its low-curvature limit reproduces the Einstein--Hilbert term, while higher curvature orders retain information not captured by the Ricci scalar alone~\cite{Bianconi:2024aju}. In particular, even on a Ricci-flat metric, the Riemann curvature can act nontrivially in the two-form sector. Earlier investigations have considered Schwarzschild-related entropy, spherically symmetric black holes, extensions using generalized entropy and the cosmological constant $\Lambda$, and vacuum perturbations in GfE \cite{Bianconi:2025rnd,Thattarampilly:2026wsw,Gohar:2026miz,Bianconi:2026smn,lxlj-ppwh,Pereira:2026aeu}. There is also an independent motivation for considering BH in GfE and treating the matter sector carefully. Observations of gravitational waves and their electromagnetic counterparts offer powerful constraints on the propagation of gravitational disturbances \cite{LIGOScientific:2017vwq,LIGOScientific:2017zic}. In GfE, a recent analysis of the tensor principal symbol identified an obstruction to real high-frequency characteristics on generic FLRW backgrounds under specified isotropy and constitutive assumptions \cite{Pereira:2026aeu}. This is a question of hyperbolicity and causal propagation, not a black-hole no-hair theorem. Nevertheless, it underscores why an admissible background solution and well-behaved perturbations must be assessed separately: establishing the absence of a static scalar profile does not, on its own, resolve the propagation properties of the full gravitational theory.

In this work, we examine the simplest bosonic truncation, $\lvert\Phi\rangle=\phi\oplus0\oplus0$, allowing $\phi$ to be complex but requiring the field itself to be static and spherically symmetric. We first reduce the \emph{full} form-space matter equation, rather than truncating the action before variation. This procedure reveals a two-form condition that is invisible in a scalar-only variation. Next we derive a radial equation for arbitrary static spherical metric functions. An integration by parts leads to a nonperturbative, conditional no-hair theorem that does not require the geometry to be Schwarzschild. The proof only demands positivity of the two auxiliary-field eigenvalues actually sampled by the scalar, together with a nonnegative effective mass squared.

The curvature dependence of that theorem leads to a second question. If the GfE vacuum geometry has a region of negative Ricci scalar, can a sufficiently large positive nonminimal coupling $\xi$ support a scalar cloud? To address this, we present the weak-curvature vacuum solution at the first nonvanishing order, identify where $m^2+\xi R$ changes sign, and then prove a stronger spectral inequality for the linearized scalar equation. The result identifies a finite range in which the elementary positivity argument has ceased to apply but regular decaying zero modes remain excluded.

The paper is organized as follows. Section~\ref{sec:framework} sets out the form-space conventions and auxiliary equations. Section~\ref{sec:scalar} derives the exact pure-scalar system. Section~\ref{sec:exact} establishes the static spherical identity and no-hair theorem. Section~\ref{sec:vacuum} constructs the perturbative vacuum geometry, and Sec.~\ref{sec:bifurcation} studies the potential for curvature-induced scalar modes. We conclude in Sec.~\ref{sec:discussion}; Technical curvature formulas and independent variational checks are collected in the appendices.

\section{GfE framework and field equations}\label{sec:framework}

At each spacetime point, consider the eleven-dimensional bosonic form space
\begin{equation}
 \mathcal V_x=\Lambda^0T_x^*M\oplus\Lambda^1T_x^*M\oplus\Lambda^2T_x^*M.
 \label{eq:formspace}
\end{equation}
Its reference form-space metric and curvature tensor are represented blockwise by \cite{Bianconi:2024aju} as
\begin{align}
 \widetilde g&=1\oplus g_{\mu\nu}\oplus g^{(2)}_{\mu\nu\rho\sigma},\nonumber\\
 g^{(2)}_{\mu\nu\rho\sigma}&=\tfrac12(g_{\mu\rho}g_{\nu\sigma}-g_{\mu\sigma}g_{\nu\rho}),\nonumber\\
 \widetilde{\mathcal R}&=R\oplus R_{\mu\nu}\oplus R_{\mu\nu\rho\sigma}.
 \label{eq:referenceblocks}
\end{align}
The field $\lvert\Phi\rangle$ comprises a scalar, one-form, and two-form. Products such as $\lvert\Phi\rangle\langle\Phi\rvert$ are defined degree by degree and they preserve form degree rather than forming unrestricted outer products between the three sectors~\cite{Bianconi:2024aju}. This convention is essential for the block decomposition used below.

Writing $D=d+\delta$ for the Hodge--Dirac operator (with $\delta$ the codifferential), the matter operator and the deformed topological metric are~\cite{Bianconi:2024aju}
\begin{align}
 \widetilde M&=D\lvert\Phi\rangle\langle\Phi\rvert D^\dagger+(m^2+\xi R)\lvert\Phi\rangle\langle\Phi\rvert,\label{eq:Mdef}\\
 \widetilde{\mathbf G}&=\widetilde g+\alpha\widetilde M-\beta\widetilde{\mathcal R}.
 \label{eq:metricdeformation}
\end{align}
Here $\alpha$ and $\beta$ control the matter and curvature deformations; both coefficients are positive, and we assume $\alpha\ne0$ so that the scalar sector is dynamical. The logarithm in the GfE action acts not on the covariant form-space tensor itself, but on the mixed operator obtained by contracting with the inverse reference metric, as
\begin{equation}
 \Theta=\widetilde{\mathbf G}\,\widetilde g^{-1}.
 \label{eq:ThetaDef}
\end{equation}

Note that raising an index with $\widetilde g^{-1}$ is not the same operation as inverting the resulting matrix. The foundational trace-log action can be written as
\begin{equation}
 S[g,\Phi]= -\mpl^{-4}\int\dd^4x\sqrt{-g}\,\TrF\ln\Theta.
 \label{eq:action}
\end{equation}
The real analytic branch used here is connected continuously to $\Theta=\Id$, where the logarithm is unambiguous~\cite{lxlj-ppwh}. Introducing now an independent mixed operator $\widehat\Theta$ and a Lagrange multiplier $\mathcal G$ gives an auxiliary description wherever the inverse exists \cite{Bianconi:2024aju}:
\begin{equation}
 \mathcal L_{\rm aux}=-\TrF\ln\widehat\Theta
 -\TrF\bigl[\mathcal G(\Theta-\widehat\Theta)\bigr].
\end{equation}
Variation with respect to $\mathcal G$ fixes $\widehat\Theta=\Theta$, while variation with respect to $\widehat\Theta$ gives $\mathcal G=\widehat\Theta^{-1}$. Thus its algebraic constraint is given by
\begin{equation}
 \mathcal G^{-1}=\Theta
 =\Id+\alpha\widetilde M\widetilde g^{-1}
       -\beta\widetilde{\mathcal R}\widetilde g^{-1},
 \label{eq:Gconstraint}
\end{equation}
with $\widetilde g_{\mathcal G}=\mathcal G^{-1}\widetilde g$. The bosonic matter equation in turn takes the form
\begin{equation}
 D\widetilde g_{\mathcal G}^{-1}D\lvert\Phi\rangle
 +\widetilde g_{\mathcal G}^{-1}(m^2+\xi R)\lvert\Phi\rangle=0.
 \label{eq:matterEOM}
\end{equation}
Finally, the associated corrected metric equation may be summarized as
\begin{equation}
 R^{\mathcal G}_{(\mu\nu)}-\frac12g_{\mu\nu}
 (R_{\mathcal G}-2\Lambda_{\mathcal G})+D_{(\mu\nu)}
 =\frac{\alpha}{\beta}T_{(\mu\nu)},
 \label{eq:metricEOM}
\end{equation}
where
\begin{equation}
 \Lambda_{\mathcal G}=\frac{1}{2\beta}\TrF
 \bigl(\mathcal G-\Id-\ln\mathcal G\bigr).
\end{equation}

The curvature and derivative terms on the left side are those of the corrected auxiliary formulation \cite{Bianconi:2024aju}. The scalar-sector theorem below requires Eq.~\eqref{eq:Gconstraint} and the full Eq.~\eqref{eq:matterEOM}; it does not assume an explicit solution of Eq.~\eqref{eq:metricEOM}.

\section{Exact pure-scalar reduction}\label{sec:scalar}

Having the dynamics of the theory established, we now specialize the matter sector to the description given by:
\begin{equation}
 \lvert\Phi\rangle=\phi\oplus0\oplus0,
 \label{eq:purescalar}
\end{equation}
with $\phi$ being possibly complex. Because $\delta\phi=0$, the first Hodge--Dirac action produces only a one-form,
\begin{equation}
 D\lvert\Phi\rangle=0\oplus d\phi\oplus0.
\end{equation}
Consequently the three matter blocks are
\begin{align}
 M_{(0)}&=U|\phi|^2,\qquad U\equiv m^2+\xi R,\nonumber\\
 [M_{(1)}]_{\mu\nu}&=(\nabla_\mu\bar\phi)(\nabla_\nu\phi),\nonumber\\
 M_{(2)}&=0.
 \label{eq:Mblocks}
\end{align}
In particular, the field's gradient deforms the one-form block, while the potential term deforms the scalar block. No scalar bilinear appears directly in the two-form block.

Introduce the curvature-only factors
\begin{equation}
 A_0=1-\beta R,\qquad [A_1]_\mu{}^\nu
 =\delta_\mu{}^\nu-\beta R_\mu{}^\nu.
 \label{eq:Ablocks}
\end{equation}
The algebraic constraint \eqref{eq:Gconstraint} is then exactly
\begin{align}
 \mathcal G_{(0)}^{-1}&=A_0+\alpha U|\phi|^2,\label{eq:G0}\\
 [\mathcal G_{(1)}^{-1}]_\mu{}^\nu
   &=[A_1]_\mu{}^\nu
     +\alpha(\nabla_\mu\bar\phi)(\nabla^\nu\phi),\label{eq:G1}\\
 \mathcal G_{(2)}^{-1}&=I_{(2)}-\beta\mathbb R_{(2)}.
 \label{eq:G2}
\end{align}
The full curvature operator $\mathbb R_{(2)}$ acts on two-forms. Its omission from the pure-scalar matter equation below is a consequence of form degree, not an assumption that the Riemann tensor vanishes.

Defining the dressed one-form current as
\begin{equation}
 J=\widetilde g_{\mathcal G,(1)}^{-1}d\phi,
 \qquad J_\mu=[\mathcal G_{(1)}]_\mu{}^\nu\nabla_\nu\phi.
 \label{eq:Jdef}
\end{equation}
by acting with $D$ once more yields a scalar and a two-form: $DJ=\delta J\oplus0\oplus dJ$, with $\delta J=-\nabla_\mu J^\mu$. Meanwhile, the potential term occupies only the zero-form sector. Thus the \emph{full} matter equation gives two independent conditions,
\begin{equation}
 {\quad\nabla_\mu J^\mu-\frac{U\phi}{A_0+\alpha U|\phi|^2}=0,
 \qquad dJ=0.\quad}
 \label{eq:scalarconsistency}
\end{equation}
The second equation means that the dressed current must be closed. It would not follow from varying a scalar-only truncation before checking the other form degrees.

For completeness, Eq.~\eqref{eq:G1} can be inverted without writing all components explicitly. Suppose $A_1$ is invertible, and set
\begin{align}
 X_\beta&=(\nabla^\mu\phi)[A_1^{-1}]_\mu{}^\nu\nabla_\nu\bar\phi,\nonumber\\
 Y_\beta&=(\nabla^\mu\phi)[A_1^{-1}]_\mu{}^\nu\nabla_\nu\phi.
\end{align}
The rank-one inverse formula gives
\begin{equation}
 J_\mu=[A_1^{-1}]_\mu{}^\nu\nabla_\nu\phi
 -\frac{\alpha Y_\beta}{1+\alpha X_\beta}
  [A_1^{-1}]_\mu{}^\nu\nabla_\nu\bar\phi,
 \label{eq:Jgeneral}
\end{equation}
provided $1+\alpha X_\beta\ne0$. This expression is exact and is useful outside spherical symmetry, where $dJ=0$ imposes a genuine additional restriction.

\subsection{The Ricci flat case}

If $R_{\mu\nu}=0$, then $R=0$ and $A_0=A_1=1$. The sampled blocks become
\begin{equation}
 \mathcal G_{(0)}^{-1}=1+\alpha m^2|\phi|^2,
 \qquad \mathcal G_{(1)}^{-1}=I_{(1)}+\alpha\nabla\bar\phi\otimes\nabla\phi,
\end{equation}
while $\mathcal G_{(2)}^{-1}=I_{(2)}-\beta\mathbb R_{(2)}$ generally remains nontrivial. Hence the pure-scalar equation on a Ricci-flat geometry is independent of \emph{explicit} $\beta$ dependence even if the Weyl tensor is nonzero. In this special case,
\begin{equation}
 J_\mu=\nabla_\mu\phi
 -\frac{\alpha(\nabla\phi)^2}{1+\alpha\nabla_\rho\phi\nabla^\rho\bar\phi}
   \nabla_\mu\bar\phi,
 \label{eq:Jricciflat}
\end{equation}
where $(\nabla\phi)^2=\nabla_\rho\phi\nabla^\rho\phi$. For a real scalar with null gradient this current reduces to $d\phi$; this special kinematics is outside a nonconstant static radial profile, whose gradient has positive norm in the exterior. It is important to state and understand that evaluating a matter equation on Schwarzschild is \emph{not} a proof that Schwarzschild solves the full finite-coupling GfE metric equation.
The vacuum geometry is constructed in Sec.~\ref{sec:vacuum}.

\section{Static spherical black holes: an exact no-hair theorem}\label{sec:exact}

We will now consider the general areal-radius ansatz
\begin{equation}
 \dd s^2=-A(r)\dd t^2+B(r)\dd r^2+r^2\dd\Omega^2,
 \qquad \phi=\phi(r),
 \label{eq:sss}
\end{equation}
where $A,B>0$ outside the outer event horizon. The mixed Ricci tensor is diagonal by symmetry. The only eigenvalues in Eqs.~\eqref{eq:G0}--\eqref{eq:G1} sampled by a radial scalar are
\begin{align}
 \Theta_0&\equiv1-\beta R+\alpha(m^2+\xi R)|\phi|^2,
 \label{eq:Theta0}\\
 \Theta_r&\equiv1-\beta R^r{}_r+\alpha B^{-1}|\phi'|^2.
 \label{eq:Thetar}
\end{align}
The curvature invariants $R$ and $R^r{}_r$ in these expressions are calculated from the general metric, not from a Schwarzschild approximation; explicit formulas are given in Appendix~\ref{app:curvature}.

By symmetry, the only nonzero component of the current is
\begin{equation}
 J=\frac{\phi'}{\Theta_r}\,\dd r,
 \qquad J^r=\frac{\phi'}{B\Theta_r}.
 \label{eq:Jradial}
\end{equation}
Accordingly $dJ=0$ identically, so the pure-scalar ansatz is consistent with the extra form-degree equation. Since $\sqrt{-g}=r^2\sin\theta\sqrt{AB}$, the remaining field equation becomes
\begin{equation}
 {\;
 \frac{1}{r^2\sqrt{AB}}\frac{\dd}{\dd r}
 \left[\frac{r^2\sqrt{A/B}\,\phi'}{\Theta_r}\right]
 -\frac{(m^2+\xi R)\phi}{\Theta_0}=0.\;}
 \label{eq:radial}
\end{equation}
No expansion in $\alpha$, $\beta$, or the scalar amplitude was used, so this equation is exact. The two-form curvature block and the remaining one-form eigenvalues still enter the independent metric equations, but not this radial matter equation. Appendix~\ref{app:variation} confirms Eq.~\eqref{eq:radial} by direct variation of the reduced trace-log action.

\subsection{Integrated identity and its boundary conditions}

Multiplying Eq.~\eqref{eq:radial} by $\bar\phi\,r^2\sqrt{AB}$, integrating from $r_h$ to infinity, and integrating by parts gives
\begin{align}
 &\left[\frac{r^2\sqrt{A/B}\,\bar\phi\phi'}{\Theta_r}
       \right]_{r_h}^{\infty}\nonumber\\
 &\quad-\int_{r_h}^{\infty}\!\dd r\,r^2\sqrt{AB}
 \left[\frac{|\phi'|^2}{B\Theta_r}
       +\frac{(m^2+\xi R)|\phi|^2}{\Theta_0}\right]=0.
 \label{eq:boundaryidentity}
\end{align}
For a regular nonextremal horizon, let $x=r-r_h$ and suppose
\begin{align}
 A&=a_1x+\order(x^2),\nonumber\\
 B&=b_{-1}x^{-1}+\order(1),\qquad a_1,b_{-1}>0.
\end{align}
Then $\sqrt{A/B}=\order(x)$. If $\phi$, $\phi'$, and $\Theta_r^{-1}$ are finite in this regular radial description, the horizon boundary term vanishes. We will also require an asymptotically flat exterior with a sufficiently rapidly decaying field, so that $r^2\sqrt{A/B}\,\bar\phi\phi'/\Theta_r\to0$. Under these assumptions Eq.~\eqref{eq:boundaryidentity} reduces to
\begin{equation}
 {\quad
 \int_{r_h}^{\infty}\!\dd r\,r^2\sqrt{AB}
 \left[\frac{|\phi'|^2}{B\Theta_r}
 +\frac{(m^2+\xi R)|\phi|^2}{\Theta_0}\right]=0.
 \quad}
 \label{eq:nohairidentity}
\end{equation}

\begin{theorem}[Conditional pure-scalar no-hair theorem]
\label{thm:nohair}
Let Eq.~\eqref{eq:sss} describe a regular, asymptotically flat static spherical black-hole exterior, and let a pure scalar $\phi(r)$ solve the full GfE matter equation. Assume the boundary conditions used in Eq.~\eqref{eq:nohairidentity} and, everywhere outside the horizon,
\begin{equation}
 \Theta_0>0,\qquad \Theta_r>0,\qquad m^2+\xi R\geq0.
 \label{eq:signconditions}
\end{equation}
Then any scalar field satisfying $\phi\to0$ at infinity vanishes identically: $\phi(r)=0$.
\end{theorem}
\begin{proof}
For $A,B>0$ and the stated eigenvalue signs, both terms in the integrand of Eq.~\eqref{eq:nohairidentity} are nonnegative. Its vanishing integral forces $|\phi'|^2=0$ throughout the exterior. The scalar is constant, and the asymptotic boundary condition makes that constant zero.
\end{proof}

The result shown applies to complex as well as real static scalars. In particular, $\xi=0$ and $m^2\geq0$ make the third condition automatic and neither Schwarzschild geometry nor perturbative backreaction is assumed. On the other hand, Eq.~\eqref{eq:signconditions} is a substantive branch condition: it cannot simply be inferred from weak coupling if a proposed solution crosses a zero of $\Theta_0$ or $\Theta_r$. A synchronized field $\phi(t,r)=e^{-\iu\omega t}\psi(r)$, a singular horizon configuration, or a field with different asymptotics is outside the theorem's assumptions.

\section{The perturbative GfE vacuum black hole}\label{sec:vacuum}

The theorem is independent of a particular vacuum solution, but the question of curvature-induced scalarization is not. We therefore examine the Ricci scalar of the static GfE vacuum near the general-relativistic limit. We start by setting $\phi=0$ and defining the curvature endomorphism $\mathcal R=\widetilde{\mathcal R}\widetilde g^{-1}$. The trace-log expansion is given by
\begin{align}
 \TrF\ln(\Id-\beta\mathcal R)
 &=-\beta Q_1-\frac{\beta^2}{2}Q_2
   -\frac{\beta^3}{3}Q_3+\order(\beta^4),\nonumber\\
 Q_n&\equiv\TrF(\mathcal R^n).
 \label{eq:Wexp}
\end{align}
With the two-form normalization in Eq.~\eqref{eq:referenceblocks}, the first two traces are
\begin{align}
 Q_1&=3R,\label{eq:Q1}\\
 Q_2&=R^2+R_{\mu\nu}R^{\mu\nu}
       +R_{\mu\nu\rho\sigma}R^{\mu\nu\rho\sigma}
     =\mathcal E_4+5R_{\mu\nu}R^{\mu\nu},
 \label{eq:Q2}
\end{align}
where $\mathcal E_4=R^2-4R_{\mu\nu}R^{\mu\nu}+R_{\mu\nu\rho\sigma}R^{\mu\nu\rho\sigma}$ is the four-dimensional Euler (Gauss--Bonnet) density. Its integral does not modify the local bulk equations in four dimensions, while the variation of $R_{\mu\nu}R^{\mu\nu}$ vanishes when evaluated on a Ricci-flat background. This explains why a Ricci-flat Einstein solution receives no local source through quadratic curvature order, even though the GfE action contains nontrivial higher-curvature operators \cite{Lovelock:1971yv,Stelle:1976gc,lxlj-ppwh}.

By contrast,
\begin{equation}
 Q_3=R^3+R^\mu{}_\nu R^\nu{}_\rho R^\rho{}_\mu
     +\operatorname{Tr}_{\Lambda^2}(\mathcal R_{(2)}^3)
 \label{eq:Q3}
\end{equation}
contains a non-topological Riemann-cubic contribution. Its variation can source the Schwarzschild background. Because the leading Einstein term in the action is proportional to $\beta$, the order-$\beta^3$ source generates a metric response at relative order $\beta^2$. This statement concerns the perturbative analytic branch; it is not a claim that Schwarzschild solves the finite-$\beta$ theory exactly.

For the metric variation we keep three functions in the spherical ansatz,
\begin{equation}
 \dd s^2=-e^{2u(r)}\dd t^2+e^{2v(r)}\dd r^2
             +\rho(r)^2\dd\Omega^2,
 \label{eq:threefunction}
\end{equation}
and impose the areal-radius gauge $\rho=r$ only after variation. This avoids discarding the angular field equation by prematurely fixing the radial coordinate. In spherical symmetry the two-form curvature operator is diagonal in the basis of coordinate two-planes; its eigenvalues are $2K_{ab}$, with $K_{ab}=R_{abab}/(g_{aa}g_{bb})$ (no sum). The eleven-dimensional trace thus reduces to a sum of logarithms of the scalar, one-form, and six two-form eigenvalues.

\subsection{Metric functions at leading nontrivial order}

Write $f=1-2M/r$, with $M$ the asymptotic mass, and parameterize
\begin{equation}
 A=f(1+\beta^2F)+\cdots,\qquad
 B=f^{-1}(1+\beta^2G)+\cdots .
 \label{eq:metricFG}
\end{equation}
The equations obtained from the cubic curvature source can be organized as
\begin{align}
 \frac{\dd}{\dd r}\big[(r-2M)G\big]
 &=\frac{120M^2}{r^6}-\frac{784M^3}{3r^7},
 \label{eq:Geq}\\
 F+G&=-\frac{24M^2}{r^6}.
 \label{eq:FplusG}
\end{align}
A detailed integration, including the mass and time-normalization constants, appears in Appendix~\ref{app:vacuum}. Fixing $M$ to be the ADM mass and normalizing the asymptotic time coordinate gives
\begin{align}
 F(r)&=\frac{40M^3}{9r^6(r-2M)},\label{eq:Fsol}\\
 G(r)&=\frac{8M^2(49M-27r)}{9r^6(r-2M)}.
 \label{eq:Gsol}
\end{align}
Although these functions have poles at the \emph{uncorrected} Schwarzschild radius, the perturbative expressions for $A$ and $B^{-1}$ are finite there:
\begin{align}
 A(r)&=1-\frac{2M}{r}
    +\frac{40\beta^2M^3}{9r^7}+\cdots,
 \label{eq:Afinal}\\
 B(r)^{-1}&=1-\frac{2M}{r}
    +\frac{24\beta^2M^2}{r^6}
    -\frac{392\beta^2M^3}{9r^7}+\cdots.
 \label{eq:Bfinal}
\end{align}
Here the dots denote terms beyond the displayed order, including contributions that need not be pure even powers of $\beta$. Away from $r=2M$,
\begin{equation}
 AB=1-\frac{24\beta^2M^2}{r^6}+\cdots.
 \label{eq:ABfinal}
\end{equation}
The metric is therefore not of the Schwarzschild form $AB=1$. The $tt$ and $rr$ equations describe an effective anisotropic geometric source, not an additional matter fluid. For example, if its Einstein-tensor representation is written as $G^\mu{}_\nu=\kappa_{\rm eff}T^\mu{}_{\nu,\rm ent}$, then
\begin{align}
 G^r{}_r-G^t{}_t
 &=\frac{1}{Br}\frac{\dd}{\dd r}\ln(AB)\nonumber\\
 &=\frac{144\beta^2M^2}{r^8}\left(1-\frac{2M}{r}\right)+\cdots.
 \label{eq:EinsteinDifference}
\end{align}
The nonzero difference measures a radial effective null-energy combination, while the angular component differs from the radial pressure. This interpretation is bookkeeping for the higher-curvature terms and does not introduce an independent fluid model.

The common outer zero of the two perturbative horizon functions is shifted to
\begin{equation}
 r_h=2M-\frac{5\beta^2}{72M^3}+\cdots .
 \label{eq:horizon}
\end{equation}
To the order needed below, the corrected curvature is
\begin{align}
 R(r)&=\frac{16\beta^2M^2(45r-106M)}{3r^9}+\cdots,
 \label{eq:Rvac}\\
 R^r{}_r(r)&=\frac{16\beta^2M^2(27r-58M)}{3r^9}+\cdots.
 \label{eq:Rrrvac}
\end{align}
Unlike Schwarzschild, the corrected vacuum is not Ricci flat. The perturbative regime requires $|\beta|/M^2\ll1$; extrapolating these formulas to small masses or to singular auxiliary-field branches is not justified.

\section{Nonminimal coupling and the absence of scalar zero modes}\label{sec:bifurcation}

\subsection{Where the elementary sign condition fails}

On the corrected vacuum, $\Theta_0=1-\beta R$ and $\Theta_r=1-\beta R^r{}_r$ are $1+\order(\beta^3/M^6)$ and remain positive in the perturbative exterior. For a hypothetical nonzero scalar, positive $\alpha$ and $U=m^2+\xi R\geq0$ contribute nonnegative terms to both sampled eigenvalues, so the exact theorem remains applicable as long as the coupled geometry stays on that branch. This observation is conditional: it does not establish global positivity on an arbitrary strongly backreacted solution.

The leading Ricci scalar in Eq.~\eqref{eq:Rvac} changes sign at
\begin{equation}
 r_0=\frac{106}{45}M\simeq2.356M,
 \label{eq:Rzero}
\end{equation}
with $R(2M)=-\beta^2/(6M^6)$. It is negative in the near-horizon interval $2M<r<r_0$, to leading perturbative order. For $\xi>0$, this permits $U$ to be negative near the horizon when approximately
\begin{equation}
 \xi>\frac{6m^2M^6}{\beta^2}.
 \label{eq:signthreshold}
\end{equation}
Equation~\eqref{eq:signthreshold} identifies where the \emph{sufficient} hypothesis of Theorem~\ref{thm:nohair} ceases to hold. It is not an instability criterion, nor does it prove the existence of a nonlinear scalarized black hole. In particular, a connected hairy family would have to possess a regular static zero mode in the linearization about the vacuum solution.

\subsection{A stronger non-bifurcation bound}

At linear order in the scalar amplitude, the $\phi$-dependent terms in $\Theta_0$ and $\Theta_r$ drop out. In the leading small-$|\beta|/M^2$ approximation, we retain the curvature potential $\xi R$ from Eq.~\eqref{eq:Rvac}, while replacing $A$ and $B$ by the Schwarzschild functions and the two eigenvalues by unity. This is a controlled leading-order comparison when the dimensionless strength $\xi\beta^2/M^4$ is held fixed: the curvature potential is retained even when $\xi$ is large. The resulting linear equation is
\begin{equation}
 \frac{\dd}{\dd r}\left[r(r-2M)\phi'\right]
   =r^2(m^2+\xi R)\phi,
 \label{eq:linearscalar}
\end{equation}
where $R$ is the explicitly displayed leading-order expression in Eq.~\eqref{eq:Rvac}.

\begin{proposition}[Leading-order non-bifurcation interval]
\label{prop:nobifurcation}
Suppose $m^2\geq0$, $\xi\geq0$, and $\beta\ne0$. If
\begin{equation}
 0\leq\xi\leq
 \frac{6m^2M^6}{\beta^2}+\frac{3M^4}{2\beta^2},
 \label{eq:nonbifurcationbound}
\end{equation}
then Eq.~\eqref{eq:linearscalar} admits no nonzero static solution that is regular at $r=2M$ and decays sufficiently rapidly at infinity. Consequently, at this order there is no scalarized branch bifurcating from the vacuum within this interval.
\end{proposition}

\begin{proof}
Multiply Eq.~\eqref{eq:linearscalar} by $\bar\phi$, integrate by parts, and take the real part. The regular and decaying boundary conditions imply
\begin{equation}
 \int_{2M}^{\infty}\!\dd r\,
 \left[r(r-2M)|\phi'|^2+r^2(m^2+\xi R)|\phi|^2\right]=0.
 \label{eq:linearidentity}
\end{equation}
A completion of the square provides more information than the direct sign test. For any complex $\phi$ satisfying the same boundary conditions,
\begin{align}
 \int_{2M}^{\infty}\!\dd r\,r(r-2M)|\phi'|^2
 &=\int_{2M}^{\infty}\!\dd r\,r(r-2M)
       \left|\phi'+\frac{\phi}{r}\right|^2\nonumber\\
 &\quad+\int_{2M}^{\infty}\!\dd r\,\frac{2M}{r}|\phi|^2.
 \label{eq:squareidentity}
\end{align}
Indeed, the cross term is $(r-2M)(|\phi|^2)'$, and its integral is $-\int|\phi|^2\dd r$ because $[(r-2M)|\phi|^2]_{2M}^{\infty}=0$. Substituting Eq.~\eqref{eq:squareidentity} into Eq.~\eqref{eq:linearidentity},
\begin{align}
 0={}&\int_{2M}^{\infty}\!\dd r\,r(r-2M)
 \left|\phi'+\frac{\phi}{r}\right|^2\nonumber\\
 &+\int_{2M}^{\infty}\!\dd r\,r^2
 \left[\frac{2M}{r^3}+m^2+\xi R\right]|\phi|^2.
 \label{eq:strengthenedidentity}
\end{align}
For $r\geq r_0$ the bracket is strictly positive. In $2M<r<r_0$, where $R<0$, nonnegativity of the bracket follows from
\begin{equation}
 \xi\beta^2\leq
 \frac{3\left(2Mr^6+m^2r^9\right)}
      {16M^2(106M-45r)}.
 \label{eq:pointwisebound}
\end{equation}
The right-hand side increases with $r$ throughout this interval. Its limiting value as $r\downarrow2M$ is $3M^4/2+6m^2M^6$. Under Eq.~\eqref{eq:nonbifurcationbound}, therefore, the bracket in Eq.~\eqref{eq:strengthenedidentity} is positive for every $r>2M$. Both integrals must vanish, implying $\phi=0$.
\end{proof}

In the nonempty interval
\begin{equation}
 \frac{6m^2M^6}{\beta^2}<\xi\leq
 \frac{6m^2M^6}{\beta^2}+\frac{3M^4}{2\beta^2},
 \label{eq:nontrivialinterval}
\end{equation}
the effective mass squared is negative near the horizon, yet the scalar has no regular decaying static zero mode at the order considered. For $m=0$, this becomes $0<\xi\leq3M^4/(2\beta^2)$. This is a sharper statement than the exact no-hair theorem under its sign hypotheses, but it is also more restricted: it concerns the linearized field on the perturbative vacuum background, rather than the full nonlinear coupled problem.

For $\beta$ of the order of $l_{P}^2$ and $m=0$, this interval extends to $\xi\lesssim1.5\,(M/\ell_P)^4\simeq10^{152}$ for a solar-mass black hole, so the corrected vacuum admits no static scalar hair for any physically reasonable nonminimal coupling.

\subsection{Extension of the square argument}\label{sec:generalcriterion}

The completion-of-the-square method extends to the general static spherical \emph{vacuum} background. For the linearized scalar, define
\begin{equation}
 p(r)=\frac{r^2\sqrt{A/B}}{\Theta_r^{\rm vac}},\qquad
 q(r)=\frac{r^2\sqrt{AB}}{\Theta_0^{\rm vac}}(m^2+\xi R),
 \label{eq:pqdef}
\end{equation}
with $\Theta_0^{\rm vac}=1-\beta R$ and $\Theta_r^{\rm vac}=1-\beta R^r{}_r$. Define also $s=p/r^2$. Provided $p>0$ and the boundary terms vanish, the linear equation $(p\phi')'=q\phi$ yields
\begin{align}
 0={}&\int_{r_h}^{\infty}\!\dd r\,p
   \left|\phi'+\frac{\phi}{r}\right|^2\nonumber\\
 &+\int_{r_h}^{\infty}\!\dd r\,(rs'+q)|\phi|^2.
 \label{eq:generalcriterion}
\end{align}
Thus $rs'(r)+q(r)>0$ for all $r>r_h$ is a sufficient exclusion test. In terms of the metric, this quantity equals $r\,\dd[\sqrt{A/B}/\Theta_r^{\rm vac}]/\dd r+r^2\sqrt{AB}(m^2+\xi R)/\Theta_0^{\rm vac}$. It can be applied numerically to a more accurate vacuum background without assuming that $m^2+\xi R$ itself is everywhere positive. The bounds in Proposition~\ref{prop:nobifurcation} are \emph{leading-order} boundaries: corrections to the metric and curvature alter them, and a quantitative shifted endpoint requires the higher-order background and eigenvalues. We do not claim that the displayed closed-form endpoint is an exact finite-$\beta$ threshold.

\section{Discussion and conclusions}\label{sec:discussion}

Although the form space of GfE is eleven-dimensional, a pure scalar couples to only part of it. Its derivative is a one-form and its potential a zero-form, so the scalar equation involves only $\mathcal G_{(0)}$ and $\mathcal G_{(1)}$. The two-form curvature block drops out even when it is nontrivial. The full Dirac--K\"ahler equation also imposes $dJ=0$. For a static radial field this holds automatically, but in less symmetric configurations it need not hold.

Our principal result is the exact identity \eqref{eq:nohairidentity}. It excludes regular, asymptotically decaying static scalar hair for $\Theta_0,\Theta_r>0$ and $m^2+\xi R\geq0$, irrespective of whether the GfE geometry equals Schwarzschild. It holds for complex as well as real static fields and involves no expansion in $\alpha$ or $\beta$. Minimal coupling with nonnegative mass lies within this exclusion whenever the solution remains on the indicated nonsingular branch. The result is a matter-sector theorem, not an independent construction of a fully backreacted vacuum solution.

The GfE vacuum black hole is different from Schwarzschild only at order $\beta^2$, through the cubic curvature term $Q_3$. This new GFE solution has $AB\neq1$, a horizon at $r_h=2M-5\beta^2/(72M^3)$, and a Ricci scalar that is negative just outside the horizon, for $2M<r<r_0=106M/45$.

Nonminimal coupling offers a possible way around the sign proof because the perturbative GfE vacuum has a near-horizon region with $R<0$. However, loss of positivity alone is inconclusive. The strengthened integral identity shows that, to leading order around this vacuum, a finite band of couplings with $m^2+\xi R<0$ still has no regular decaying static zero mode. Any scalarized solution beyond that band would have to be demonstrated by solving the coupled equations, and a branch disconnected from the vacuum is not addressed by the linear bound. For $\beta$ of the order of the $l_p^2$, however, the band already extends to $\xi\lesssim10^{152}$ for a solar-mass black hole, so for astrophysical black holes the exclusion covers every physically reasonable coupling, given that $\beta$ is small enough.

Limits of our results follow from the basic assumptions used to obtain it. For $\alpha>0$ the one-form dressing becomes singular for a timelike gradient with $X_\beta=-1/\alpha$, which static radial profiles cannot reach. However, time-dependent configurations are not covered by this. The theorem also assumes a regular horizon and a field that decays to zero at infinity. Synchronized fields $\phi=e^{-i\omega t}\psi(r)$, and fields with alternative asymptotics lie outside the scope of the theorem.

Several directions remain open. A full nonlinear search for nonminimally coupled scalarized solutions is needed to decide whether the band excluded by Proposition~\ref{prop:nobifurcation} is followed by genuine hairy black holes. Time-dependent and synchronized complex fields require a separate analysis. The role of non-radial form sectors needs to be studies in detail. Since Kerr is also Ricci flat, it is also corrected at order $\beta^2$ through $Q_3$; computing this correction and extending the integral identity to stationary axisymmetric backgrounds is a natural follow-up to this study. Finally, perturbative stability and causal propagation of any proposed background must be established. In view of the separate gravitational-wave hyperbolicity questions identified in Ref.~\cite{Pereira:2026aeu}, obtaining a regular static black-hole geometry is a necessary but not sufficient condition for physical viability.

\begin{acknowledgments}
The authors acknowledge the use of ChatGPT (OpenAI) and Claude (Anthropic) for assistance with literature review and language refinement during manuscript preparation. DSP acknowledges that this work was supported by Fundação para a Ciência e a Tecnologia (FCT) through national funds under the research grant UID/04434/2025 (DOI 10.54499/UID/04434/2025). VK is supported by the Kempe Foundation grant JCSMK24-507.
\end{acknowledgments}

\bibliographystyle{apsrev4-2}
\bibliography{biblio2}

\appendix
\section{Curvature of a general static spherical metric}\label{app:curvature}

We use the Riemann convention
\begin{equation}
 R^\rho{}_{\sigma\mu\nu}
 =\partial_\mu\Gamma^\rho{}_{\nu\sigma}
  -\partial_\nu\Gamma^\rho{}_{\mu\sigma}
  +\Gamma^\rho{}_{\mu\lambda}\Gamma^\lambda{}_{\nu\sigma}
  -\Gamma^\rho{}_{\nu\lambda}\Gamma^\lambda{}_{\mu\sigma}.
\end{equation}
For Eq.~\eqref{eq:sss}, the nonzero independent covariant Riemann components can be represented as
\begin{align}
 R_{trtr}&=\frac{A''}{2}-\frac{(A')^2}{4A}-\frac{A'B'}{4B},\nonumber\\
 R_{t\theta t\theta}&=\frac{rA'}{2B},\nonumber\\
 R_{r\theta r\theta}&=\frac{rB'}{2B},\nonumber\\
 R_{\theta\varphi\theta\varphi}&=r^2\sin^2\theta\left(1-\frac1B\right).
 \label{eq:Riemanncomponents}
\end{align}
The corresponding $\varphi$ components follow by factors of $\sin^2\theta$ and the usual index symmetries \cite{PopeGravPhys}. The Ricci components required for the scalar equation are
\begin{align}
 R^r{}_r&=-\frac{A''}{2AB}
   +\frac{(A')^2}{4A^2B}+\frac{A'B'}{4AB^2}
   +\frac{B'}{B^2r},\label{eq:RrrAB}\\
 R&=-\frac{A''}{AB}+\frac{A'B'}{2AB^2}
   +\frac{(A')^2}{2A^2B}-\frac{2A'}{ABr}\nonumber\\
   &\quad+\frac{2B'}{B^2r}
   +\frac2{r^2}\left(1-\frac1B\right).
 \label{eq:RAB}
\end{align}
In the diagonal two-form basis $tr,t\theta,t\varphi,r\theta,r\varphi,\theta\varphi$, the curvature endomorphism has eigenvalues $2K_{ab}$, where $K_{ab}=R_{abab}/(g_{aa}g_{bb})$. These formulas specify the blocks entering the full trace-log variation, even though the radial scalar probes only Eqs.~\eqref{eq:RrrAB}--\eqref{eq:RAB}.

\section{Vacuum integration constants and geometric stresses}\label{app:vacuum}

Integrating Eq.~\eqref{eq:Geq} yields
\begin{equation}
 (r-2M)G=C_1-\frac{24M^2}{r^5}
                       +\frac{392M^3}{9r^6}.
 \label{eq:Gintegrated}
\end{equation}
The homogeneous term $G_{\rm hom}=C_1/(r-2M)$ changes $B^{-1}$ by $-\beta^2C_1/r$, which is a mass shift $M\mapsto M+\beta^2C_1/2$. Defining $M$ as the asymptotic mass thus sets $C_1=0$. Equation~\eqref{eq:FplusG} then gives $F$ in Eq.~\eqref{eq:Fsol}; its remaining additive time-rescaling freedom is removed by $A\to1$ at infinity. The angular field equation is satisfied at this order when the two independent radial equations hold, as expected from diffeomorphism invariance.

The effective geometric-source interpretation in Sec.~\ref{sec:vacuum} can be made explicit:
\begin{align}
 \kappa_{\rm eff}\rho_{\rm ent}
 &=\frac{\beta^2M^2}{r^8}\left(120-\frac{784M}{3r}\right)+\cdots,\nonumber\\
 \kappa_{\rm eff}p_{r,\rm ent}
 &=\frac{\beta^2M^2}{r^8}\left(24-\frac{80M}{3r}\right)+\cdots,\nonumber\\
 \kappa_{\rm eff}p_{t,\rm ent}
 &=\frac{\beta^2M^2}{r^8}\left(-72+\frac{496M}{3r}\right)+\cdots.
 \label{eq:entropicstresses}
\end{align}
It follows that
\begin{equation}
 \kappa_{\rm eff}(p_{t,\rm ent}-p_{r,\rm ent})
 =-\frac{96\beta^2M^2}{r^8}\left(1-\frac{2M}{r}\right)+\cdots,
\end{equation}
so the effective pressures are generically anisotropic. These quantities are computed from the Einstein tensor of the perturbative metric and are not assigned an independent equation of state.

\section{Direct reduced-action check of the scalar equation}\label{app:variation}

Before imposing any scalar equation, spherical symmetry makes the trace of the logarithm a sum over the eleven form-space eigenvalues. The radial scalar appears only in $\Theta_0$ and $\Theta_r$:
\begin{equation}
 \TrF\ln\Theta
 =\ln\Theta_0+\ln\Theta_r
  +\text{terms independent of }\phi,\bar\phi.
\end{equation}
After angular integration, the reduced action is
\begin{equation}
 S_{\rm red}=-\frac{4\pi}{\mpl^4}\int\dd t\,\dd r\,
             r^2\sqrt{AB}\,\TrF\ln\Theta.
\end{equation}
Varying with respect to $\bar\phi$ at fixed metric gives
\begin{align}
 \frac{\partial\big(r^2\sqrt{AB}\,\TrF\ln\Theta\big)}
 {\partial\bar\phi}
 &=\alpha r^2\sqrt{AB}\,\frac{U\phi}{\Theta_0},\nonumber\\
 \frac{\partial\big(r^2\sqrt{AB}\,\TrF\ln\Theta\big)}
 {\partial\bar\phi'}
 &=\alpha r^2\sqrt{A/B}\,\frac{\phi'}{\Theta_r}.
\end{align}
The Euler--Lagrange equation, after division by $\alpha\ne0$, is precisely Eq.~\eqref{eq:radial}. This supplies an independent variational check in static spherical symmetry. The scalar-only reduced action, by construction, does not separately display the two-form condition $dJ=0$; that condition was established from the full equation in Sec.~\ref{sec:scalar}.

\end{document}